\documentclass[aps,prl,twocolumn,showpacs,letterpaper,superscriptaddress]{revtex4}
\usepackage{graphicx}
\usepackage{amsmath}

\begin{document}
\title{Ab initio GW many-body effects in graphene}

\author{Paolo E. \surname{Trevisanutto}}
\affiliation{Institut N\'eel, CNRS \& UJF, Grenoble, France}
\affiliation{European Theoretical Spectroscopy Facility (ETSF), Paris, France}
\author{Christine \surname{Giorgetti}}
\affiliation{Laboratoire des Solides Irradi\'es, CNRS - CEA, \'Ecole Polytechnique, Palaiseau, France}
\affiliation{European Theoretical Spectroscopy Facility (ETSF), Paris, France}
\author{Lucia \surname{Reining}}
\affiliation{Laboratoire des Solides Irradi\'es, CNRS - CEA, \'Ecole Polytechnique, Palaiseau, France}
\affiliation{European Theoretical Spectroscopy Facility (ETSF), Paris, France}
\author{Massimo \surname{Ladisa}}
\affiliation{Istituto di Cristallografia, CNR, Bari, Italy}
\author{Valerio \surname{Olevano}}
\affiliation{Institut N\'eel, CNRS \& UJF, Grenoble, France}
\affiliation{European Theoretical Spectroscopy Facility (ETSF), Paris, France}
\date{\today}

\begin{abstract}
We present an {\it ab initio} numerical many-body GW calculation of the band plot in free-standing graphene.
We consider the full ionic and electronic structure introducing {\it e-e} interaction and correlation effects via a self-energy containing non-hermitian and dynamical terms.
With respect to the density-functional theory local-density approximation, the Fermi velocity is renormalized with an increase of 17\%, in better agreement with the experiment.
Close to the Dirac point the linear dispersion is modified by the presence of a kink, as observed in angle-resolved photoemission spectroscopy. We demonstrate that the kink is due to low-energy $\pi \to \pi^*$ single-particle excitations and to the $\pi$ plasmon.
The GW self-energy does not open the band gap.
\end{abstract}

\maketitle

%
%
The discovery of graphene by micromechanical cleavage \cite{Novoselov,Kim} and epitaxial grow \cite{Berger} has attracted tremendous interest in consideration of its unusual electronic properties.
In the tight-binding (TB) formalism, the graphene 2D honeycomb lattice structure gives rise to a semiconductor with zero band gap occurring at the K point in the Brillouin zone and a cone-like linear band-dispersion at low energy. This part is usually described by a massless Dirac (Weyl) dispersion ~\cite{CastroNetoGuinea}.
{\it Ab initio} density-functional theory (DFT) calculations \cite{CalandraMauri} confirm the TB linear dispersion picture and give an estimate of the Fermi velocity $v_{\rm F}$ lower by 15$\sim$20\% than the experimental value.
Recently, two angle-resolved photoemission spectroscopy (ARPES) experiments on graphene epitaxially grown on SiC \cite{Bostwick,Zhou,Zhoucondmat,comment,reply} raised the general interest. The first one \cite{Bostwick,comment} observed at low energy a nearly linear band dispersion with slight deviations in the form of small kinks interpreted as due to many-body electron-electron ({\it e-e}) and electron-phonon ({\it e-ph}) self-energy effects. The second one \cite{Zhou,Zhoucondmat,reply} provided a different picture, with the opening of a band gap occurring at the Dirac K point and attributed either to substrate (SiC) or to many-body self-energy effects.
A DFT calculation \cite{Koreans} seemed to confirm a substrate induced symmetry breaking, but recent STM measures \cite{CM_Pierre} provided some evidence to exclude it.
This situation calls for clarification about the role of {\it e-e} self-energy effects on the quasiparticle (QP) band plot, the Fermi velocity and the band gap opening.
Previous {\it ab initio} works have dealt with {\it e-ph} effects \cite{CalandraMauri,Park} and with {\it e-e} GW effects in graphene nanoribbons \cite{Prezzi}. There are also several non {\it ab initio} works \cite{MDM,Mishchenko,Vozmediano,Polini} which studied {\it e-e} self-energy effects in a 2D massless Dirac model.

%
%
In this work we calculate the band plot of free-standing undoped graphene introducing {\it e-e} interaction and correlation effects by an {\it ab initio} many-body GW self-energy~\cite{GW,GodbyNeeds}. 
We numerically simulate the full ionic and electronic structure of real graphene.
We take into account the full dynamical dependence and non-hermiticity of the self-energy by an accurate contour-deformation (CD) integration.
From the self-energy we then obtain the QP energies and the spectral function which can be directly compared with ARPES spectra.
We show that the GW self-energy renormalizes the Fermi velocity by 17\% such that it corrects the DFT underestimation and leads to a value of $1.12 \cdot 10^6$ ms$^{-1}$, in good agreement with the accurate magnetotransport measure of $1.1 \cdot 10^6$ ms$^{-1}$~\cite{Kim}.
Furthermore, the nearly linear DFT band dispersion is in GW considerably distorted. Close to the Dirac point the self-energy results in an unusual negative GW band gap correction and the appearance of a kink in the band plot, leading to a scenario similar to that observed in ARPES, Ref.~\cite{Bostwick}.
A comparison with the results of a GW plasmon-pole model calculation indicates that the kink is due to a coupling with the $\pi$ plasmon at $\sim 5$ eV and the low-energy $\pi \to \pi^*$ single-particle excitations (SPE) shoulder present in the energy-loss function. This provides a partial confirmation to the explanation given in Ref.~\cite{Bostwick}.
Finally, our results show that in free-standing graphene the GW self-energy does not open the band gap, in contrast to what is found in GW calculations on graphene nanoribbons \cite{Prezzi}.

\begin{figure}
  \includegraphics[clip,width=0.45\textwidth]{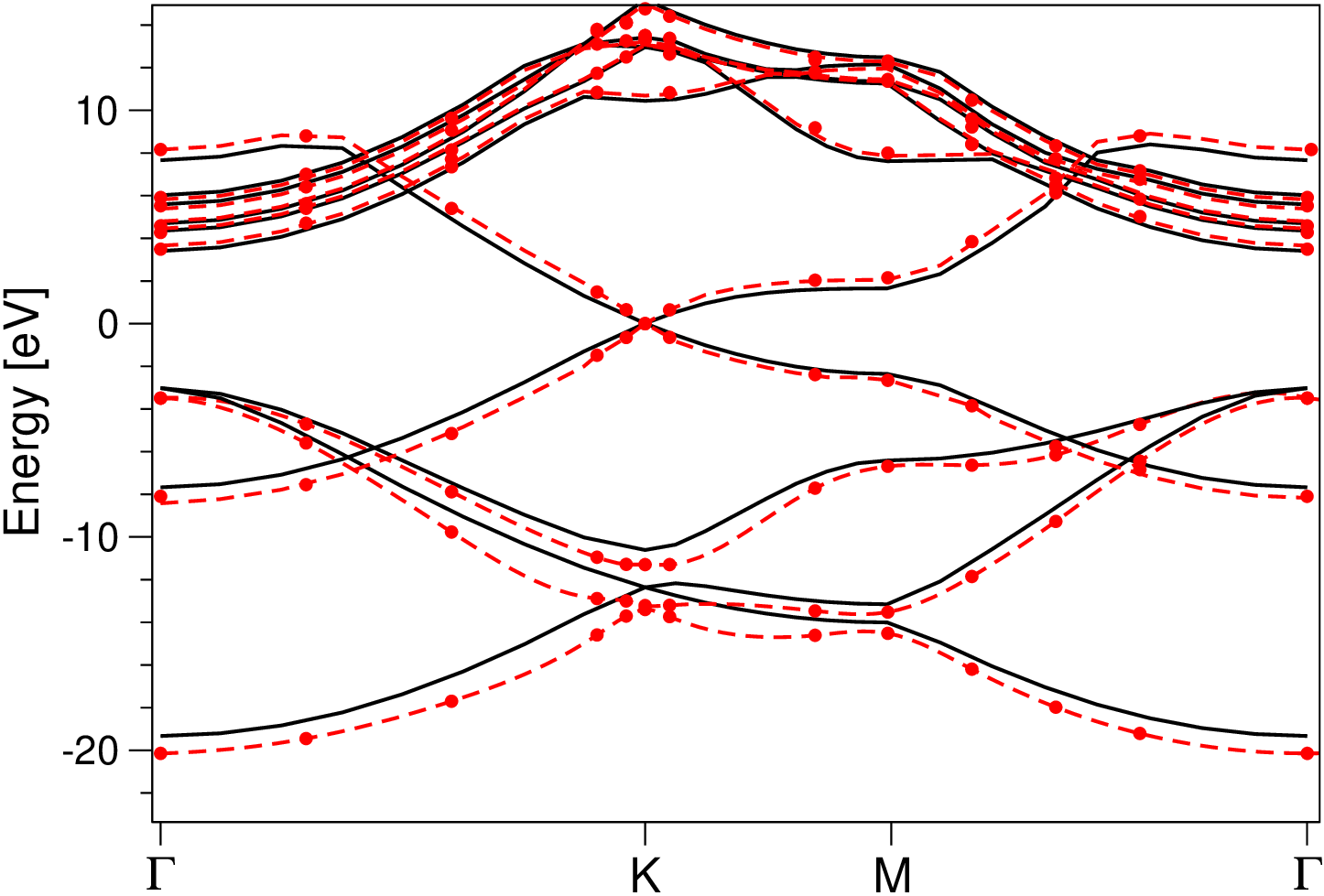}
  \caption{Band plot of graphene. Solid thick lines: DFT-LDA KS; circles and dashed lines: GW.}
  \label{fullbandplot}
\end{figure}

Our starting point is a standard ground-state DFT LDA calculation of infinite free-standing graphene.
We use a plane waves basis set (62 Ry cutoff) and periodic boundary conditions on a hexagonal cell containing 2 carbon atoms ($a=2.4$~\AA) and 38 Bohr of vacuum along the $z$ direction, large enough to isolate spurious replica of graphene sheets.
We use Martins-Trouiller norm-conserving pseudopotentials with \textit{s} and \textit{p} electrons in the valence.
We first calculate the ground-state energy and electronic density, and then the Kohn-Sham (KS) electronic structure to be used as starting point in the following GW excited-state calculation.

In the GW approximation \cite{GW,GodbyNeeds} the self-energy is 
\[
  \label{GWselfenergyomega}
  \Sigma^{\rm GW}(r,r',\omega) = \frac{i}{2 \pi}
  \int_{-\infty}^{\infty} d\omega' \,
  G(r,r',\omega-\omega')
  W(r,r',\omega')
\]
that is the product of the Green's function $G$ and the dynamically screened interaction $W(\omega)=\varepsilon^{-1}(\omega) v$ defined as the bare Coulombian interaction $v$ screened by the dynamical dielectric function $\varepsilon^{-1}(\omega)$. Vertex corrections are neglected both in the self-energy and in the polarizability (hence in $W$).
In the standard {\it ab initio} GW resolution procedure one builds $G$ and $W$ using the DFT KS electronic structure. The integral over the frequency is performed by a CD method \cite{cd} which consists in a deformation of the real axis contour such as to calculate the self-energy as an integral along the imaginary axis minus a contribution arising from the residual of the contour-included poles of $G$. This is the most accurate method to perform a GW calculation. We also considered the standard {\it plasmon-pole} model (PPM) approach \cite{GodbyNeeds}.
Once the integration is performed, we calculate the GW quasiparticle energies using a first-order perturbation theory expansion of $\Sigma$ around the LDA exchange-correlation potential $v_{xc}^{\rm LDA}$ and the KS energies $\omega=\epsilon^{\rm KS}_{nk}$,
\[
  \epsilon^{\rm GW}_{nk} = \epsilon^{\rm KS}_{nk} +
    Z \big\langle nk \big| \Sigma^{\rm GW}\big(\omega=\epsilon^{\rm KS}_{nk} \big) - v_{xc}^{\rm LDA} \big| nk \big\rangle
\]
where $Z = (1 - \left. \partial \Sigma^{\rm GW} / \partial \omega \right|_{\omega=\epsilon^{\rm KS}_{nk}} )^{-1}$ is the renormalization factor. We compare these energies to the positions of QP peaks in the spectral function
\[
  A_n(k,\omega) = \frac{\pi^{-1} | \Im \Sigma_n(k,\omega)|}
    { [\omega - \epsilon^{\rm KS}_{nk} + v^{xc}_{nk} - \Re \Sigma_n(k,\omega) ]^2
      + [\Im \Sigma_n ]^2 }
\]
calculated at the given $k$ point in the Brillouin zone and projected on the considered $n$ band. 

We used the \textsc{ABINIT} code. Convergence was achieved with 715 plane waves (10 Ry) to represent the wavefunctions and the exchange part $\Sigma_x$ of the self-energy, 150 and 200 bands to calculate respectively $W$ and $\Sigma$   \cite{EPAPS}. The dimension of $W$ and of the correlation part $\Sigma_c$ as reciprocal space matrices was 169 (6 Ry). Imaginary axis integrations were carried out by a Gauss quadrature using 10 nodes for the most accurate calculations and $W$ was sampled every 0.2 eV over 100 frequencies on the real axis. The Brillouin zone was sampled with a (10 10 1) Monkhorst-Pack k-point grid.

\begin{figure}
  \includegraphics[clip,width=0.45\textwidth]{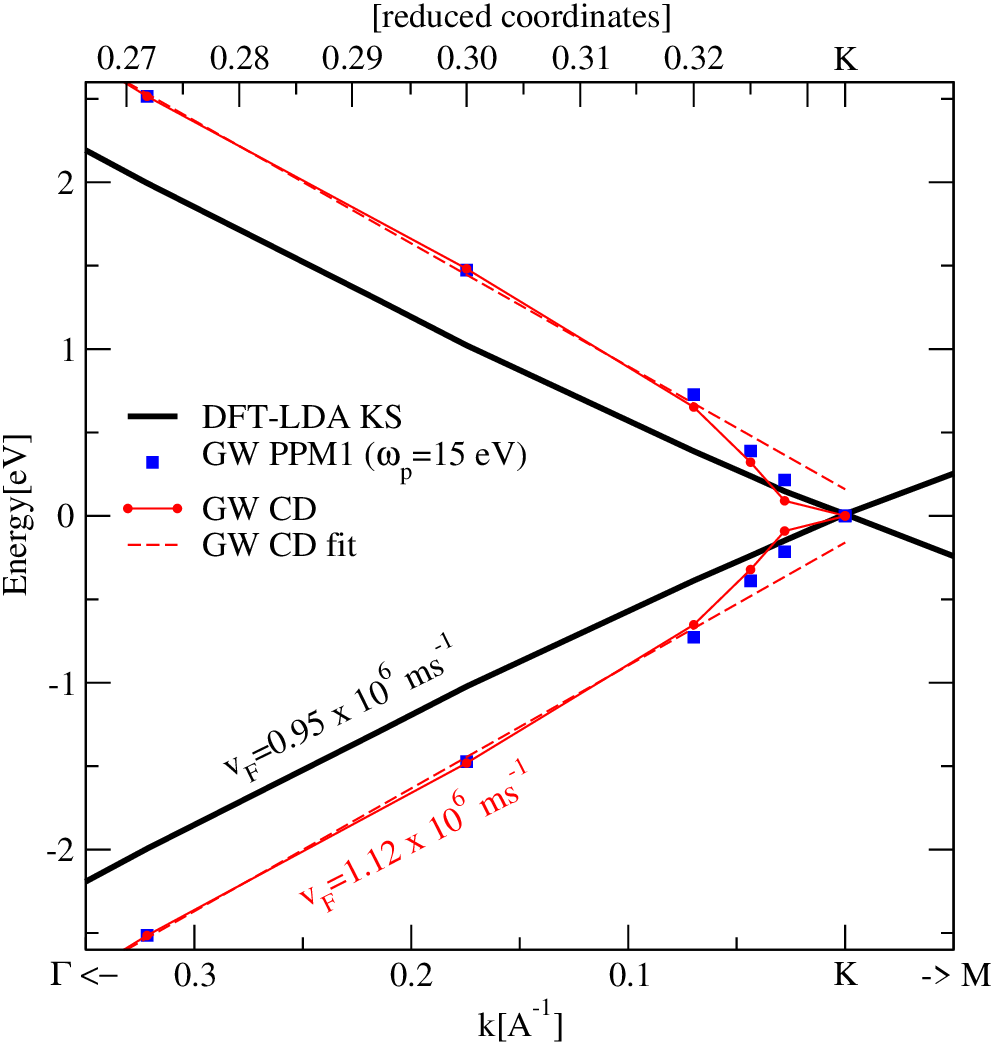}
  \caption{Focus on the band plot Dirac point region. Solid thick line: DFT-LDA KS; circles and thin lines: GW CD; squares: GW PPM1 ($\omega_p\simeq 15$ eV); dashed line: fit over the GW CD linear region.}
  \label{bandplot}
\end{figure}

%
%
In Fig.~\ref{fullbandplot} we compare the KS DFT-LDA (thick lines) and the quasiparticle GW (circles and dashed lines) electronic structures.
Exchange end correlations effects slightly affect the band shapes. Relevant effects are a lowering of the $\sigma$ bands and an increase (up to $+20$\%) of the gaps at M ($4 \to 4.8$ eV) and at $\Gamma$ ($6.4 \to 7$ eV). This is a normal behavior of GW and in agreement with other calculations \cite{Miyake,Attaccalite}.
We now focus on the Fermi energy Dirac point (K) region (Fig.~\ref{bandplot}). As previously obtained \cite{CalandraMauri}, the DFT KS $\pi$ and $\pi^*$ (thick lines) band dispersion is linear in the first $\sim 0.5$ \AA$^{-1}$ from the Dirac K point. The DFT KS Fermi velocity is $0.95 \cdot 10^6$ ms$^{-1}$. This underestimates by a 15\% the experimental value.
The dots and thin lines represent the GW band plot calculated by the CD method.
The first evident self-energy effect is the loss of linearity along the region $0.05$~\AA$^{-1}$ close to the Dirac K point. However outside that region the linearity is soon recovered but with a slope larger than the DFT KS. A fit of the GW band with a straight line (dashed line) gives a Fermi velocity of $1.12 \cdot 10^6$ ms$^{-1}$ ($1.14$ with $a_{\rm th}=2.45$~\AA). Thus the GW self-energy {\it renormalizes} by a $+17$\% the DFT Fermi velocity and achieves a good agreement with the experimental measure of $1.1 \cdot 10^6$ ms$^{-1}$ \cite{Kim}. The residual overestimation of 2$ \sim $4\% should be compensated by negative {\it e-ph} renormalization effects ($-4$\% in Ref.~\cite{Park}).

We now focus on the non-linear region.
In all the k-points far from the Dirac point, the GW correction acts in the usual direction to open the gap between DFT KS bands. On the other hand at $k = (0.328, 0.328, 0)$ (reduced coordinates), that is at $\sim 0.025$ \AA$^{-1}$ from K, we have found an unusual {\em negative GW correction} of $-0.12$ eV \footnote{This GW result is stable over four different k-meshes, increasing the convergence parameters and also moving to $k=0.329$.}
which generates a {\it kink} at $\sim 0.1$ eV from the Dirac point.
This result reproduces the experimental ARPES scenario of Ref.~\cite{Bostwick} where a kink interpreted as due to {\it e-e} many-body effects is found more or less in this position.
The position of our GW kink is also close to the position indicated in the other ARPES experiment (gray arrows in Fig.~3(d) of Ref.~\cite{Zhoucondmat} at 0.035 \AA$^{-1}$ and 0.17 eV).

\begin{table}
\begin{center}
\begin{tabular}[c]{|llccccc|}
\hline
k & method & $- \Delta v_{xc}^{\rm LDA}$ & $\Delta\Sigma_x$ & $\Delta\Sigma^{\rm GW}_c$ & $Z$ & $\Delta\epsilon^{\rm GW}$ \\
\hline
K     & CD   & $-0.023$ & $0.032$ & $-0.008$ & $0.729$ & $+0.001$ \\
0.328 & CD   & $-0.011$ & $0.493$ & $-0.689$ & $\simeq 0.71 $ & $-0.121$  \\
0.328 & PPM1 & $-0.011$ & $0.493$ & $-0.316$ & $\simeq 0.76 $ & $+0.127$ \\
0.328 & PPM2 & $-0.011$ & $0.493$ & $-0.453$ & $0.537$ & $+0.016$ \\
0.328 & CD IRcut & $-0.011$ & $0.493$ & $-0.273$ & $0.771$ & $+0.162$ \\
\hline
\end{tabular}
\end{center}
\caption{Exchange and correlation components differences between bands 5 and 4, renormalization factor $Z$ and GW band gap opening (energies in eV; Z as pure numbers).}
\label{GWtable}
\end{table}

\begin{figure}
  \includegraphics[clip,width=0.45\textwidth]{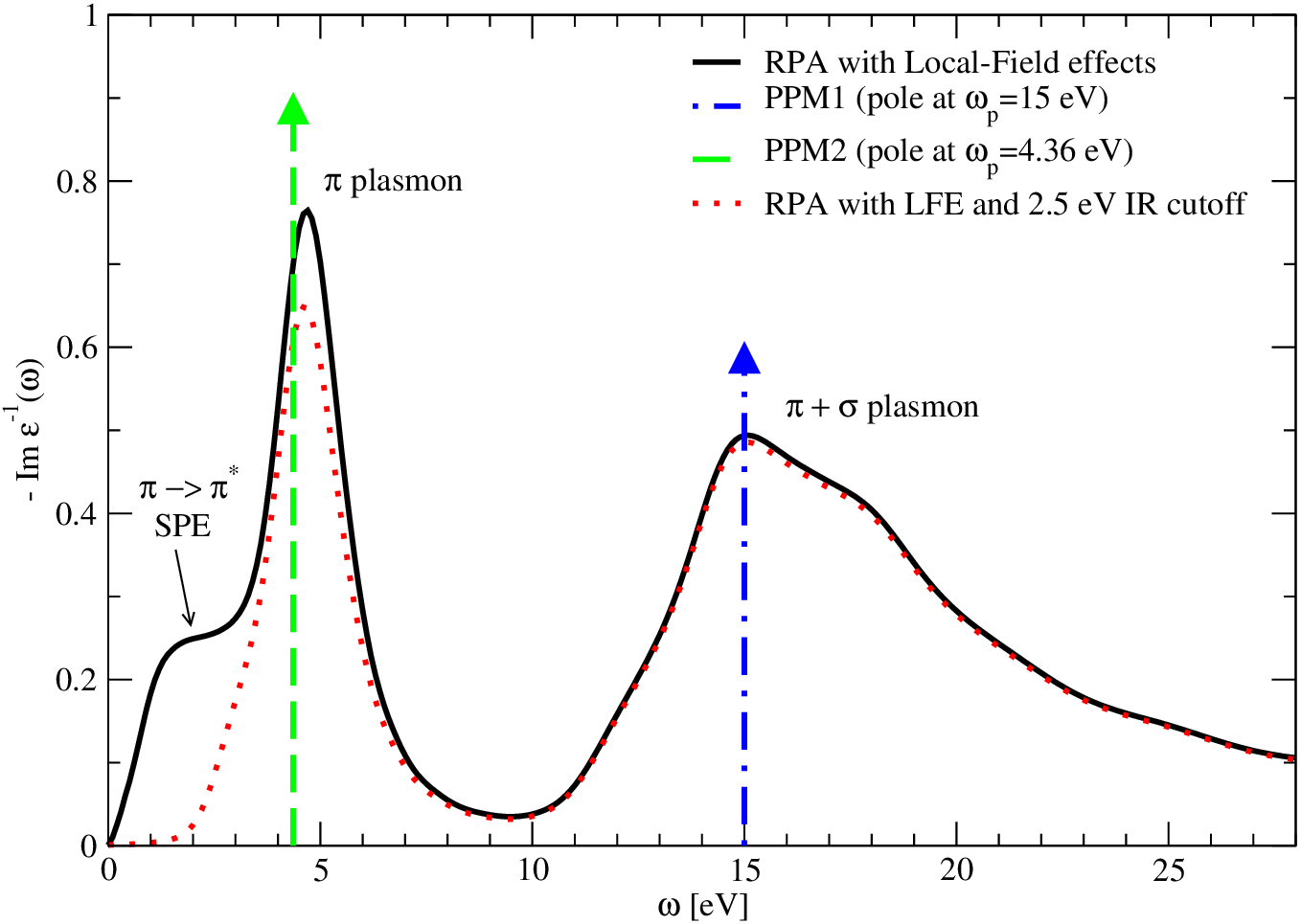}
  \caption{Graphene energy-loss function at $q \simeq 0$. Solid line: RPA with local-field effects (LFE); arrows:  positions of the PPM1 (dot-dashed) and PPM2 (dashed) poles.}
  \label{elf}
\end{figure}

The negative GW correction conjuring the kink results from an unusual balance between the exchange and the correlation energy (see Table~\ref{GWtable}). Indeed, the exchange energy difference between the bottom-of-conduction band 5 and the top-of-valence band 4, $\langle \Delta \Sigma_x \rangle$, is typically several times larger than the difference in the correlation energy $\langle \Delta \Sigma_c \rangle$ which opposes to the exchange energy \cite{GW,GodbyNeeds}. At the kink $\Delta \Sigma_c$ is $\simeq 0.7$ eV, larger than $\Delta \Sigma_x \simeq 0.5$ eV. Therefore the negative correction to the band gap is due to a correlation energy stronger than usual.
We report on Fig.~\ref{bandplot} and Table~\ref{GWtable} the result obtained by a PPM GW calculation (indicated as PPM1 and squares). Far from the Dirac point and exactly at the Dirac point (where the gap correction vanishes), the PPM GW bands precisely recover the CD result. PPM and CD start to deviate in the kink region and at the kink point the PPM provides, in contrast to CD, a positive GW correction.
This has an immediate interpretation. In the PPM approach the dynamical dependence of the energy-loss function $-\Im \varepsilon^{-1}$ entering into $W$ and the GW self-energy, is represented by a single plasmon-pole feature which is fitted to $\Re \varepsilon^{-1} = 1 + \Omega^2/(\omega^2 - \omega^2_p)$.
This is a good approximation in all systems where the energy-loss function presents a single plasmon feature.
In graphene the energy-loss function $-\Im \varepsilon^{-1}(q \simeq 0,\omega)$, as calculated from first-principles by the \textsc{DP} code in the RPA approximation (solid line in Fig.~\ref{elf}), shows two major features: the total $\pi + \sigma$ plasmon at $\sim 15$ eV and the $\pi$ plasmon at $\sim 5$ eV (see Fig. 1(e) of Ref. \cite{eels} and Ref. \cite{elf}). Furthermore, at the lowest energies there is a shoulder due to the $\pi \to \pi^*$ SPE. In our PPM1 GW calculation, the plasmon-pole frequency at $q \simeq 0$ sets to $\omega_p = 15$ eV (dot-dashed arrow) at the left side of the total plasmon peak. It reproduces the main energy-loss feature and implicitly accounts also for the low energy part of the spectrum.
Other PPM calculations are examined (PPM2 in Fig.~\ref{elf} and Table~\ref{GWtable}). Forcing the plasmon-pole to adjust to $\omega_p \simeq 5$ eV, the low energy part of the energy-loss is explicitly considered. 
In this case the resulting GW correction is around 0, closer to the correct GW CD result. Finally, we performed a CD calculation where $\varepsilon^{-1}$ is computed cutting off low energy $\pi \to \pi^*$ SPE ($2.5$ eV IR cutoff). The low energy shoulder is suppressed (dotted line in Fig.~\ref{elf}) and the intensity of the $\pi$ plasmon is also unavoidably reduced. The IR cutoff has the effect to produce a positive GW correction even beyond the PPM1 result, thus validating the good quality of the PPM. From all these results we can deduce that both the $\pi$ plasmon and the low energy $\pi \to \pi^*$ SPE shoulder provide the crucial contribution to the correlation energy and play a major role in conjuring the negative GW correction and the kink.


\begin{figure}
  \includegraphics[clip,width=0.45\textwidth]{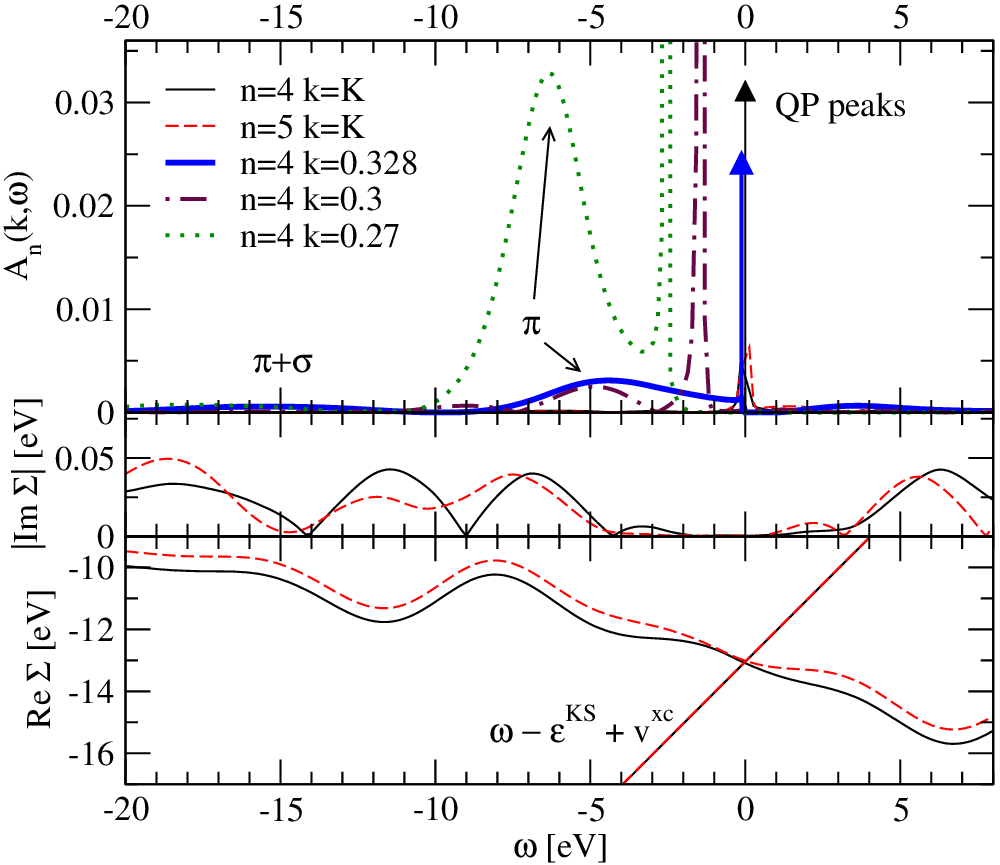}
  \caption{Graphene real and imaginary part of the self-energy and spectral function projected on bands 4 and 5 and at several k-points.}
  \label{spectralfunction}
\end{figure}

The last important result of our GW calculation is that many-body effects, within the numerical error bar, do not open the band gap at the Dirac point. Both the CD and the PPM do not change the DFT-LDA 0 band gap and band 4 and 5 keep degenerate at K.
\footnote{Hartree-Fock and Coulomb-hole plus screened-exchange approximation \cite{GW} calculations open the gap by 9 and 5 meV respectively (34 and 13 meV at self-consistence, already achieved at the 2nd iteration), all below the typical $\sim \pm 0.1$ eV numerical error of \textit{ab initio} calculations and so compatible with 0 band gap.}
We find no indication for a band gap opening even when looking at the spectral function.
In Fig.~\ref{spectralfunction} we reported the real (bottom panel) and the module of the imaginary part (middle) of the self-energy at the Dirac K point projected onto bands 4 (solid line) and 5 (dashed line).
The intersection of the real part of $\Sigma_{nk}(\omega)$ with the straight line $\omega - \epsilon^{\rm KS}_{nk} + v^{xc}_{nk}$ gives the position of the QP peak. Although $\Re \Sigma$ is different for bands 4 and 5, the intersections occur at the same point, so that the QP peaks are degenerate.
The correspondent $\Im \Sigma$ is in practice numerically 0. 
In the spectral function (top panel) we observe well defined QP peaks and weak $\pi$ and $\pi+\sigma$ plasmon satellites pointing to a normal Fermi liquid, unlike Ref.s~\cite{Vozmediano,MDM,Mishchenko,Polini}.
Thus our calculation excludes a band gap opening, even only apparent \cite{Polini}, induced by a significant transfer of weigth from QP peaks to strong satellite plasmarons \cite{MDM}.

Finally we want to discuss possible mechanisms which can explain the presented picture. A simple kink, i.e. a slope change of the bands in one point, has been obtained in Ref.~\cite{dmftkink} by invoking a purely dynamical, k-independent, mechanism; this model is most adequate for strongly correlated systems like some transition metal oxides. The situation in graphene, with its negative GW correction and consequent s-shaped kink, requires a more complex description. We can indeed understand the results by using a simplified PPM which accounts only for the low-lying $\pi \to \pi^*$ SPE and the $\pi$ plasmon. It has been shown \cite{elf} that for this plasmon we can assume a linear dispersion $\omega_p(q) = \omega_p^0 + \alpha q$ with a finite $\omega_p^0 \ne 0$. 
Together with other straightforward assumptions, this model leads to a correlation energy in the form
\[
  \Sigma_c^{\rm GW} \simeq \frac{\Omega^2}{v_F \omega_p^0} \left [\log((v_F+\alpha)k + \omega_p^0) - \log (\alpha k + \omega_p^0)\right],
\]
which in the limit $k\to \infty$ tends to a constant whereas in the limit $k \to k_F$ is linear like the exchange energy $\Sigma_x \sim k$, but with a coefficient $\Omega^2/2 {\omega_p^0}^2$ times the exchange. This hypothesis is compatible with a zero band gap at $k=k_F$ and yields a large-$k$ region where the exchange dominates and renormalizes the Fermi velocity. If $\Omega > 2 \omega_p^0$, it yields a region where the correlation dominates and provides negative GW corrections (the kink).

In conclusion, we presented an {\it ab initio} many-body GW calculation in graphene.
The GW self-energy renormalizes by a 17\% the DFT Fermi velocity but does not open a gap at K.
Close to the Dirac K point the linear band dispersion is considerably affected by correlation, leading to the appearance of a kink.

We thank W. De Heer, C. Berger, L. Magaud, D. Mayou, J.-Y. Veuillen, P. Mallet, C. Naud, P. Darancet, F. Varchon, L. Wirtz, R. Martin, T. Seyller and E. Rotenberg for useful discussions. P.E.T. was supported by ANR. Computer time has been granted by Ciment.Phynum.
We have used \textsc{ABINIT} and the \textsc{DP} code.

\end{document}